\documentclass{JHEP3}

\usepackage{amssymb,amsmath}

\newcommand{\Sa}{I}
\newcommand{\im}{\mathrm{i}}
\newcommand{\ee}{\mathrm{e}}
\newcommand{\de}{\mathrm{d}}
\DeclareMathOperator{\Hodge}{\star}
\DeclareMathOperator{\diam}{\diamond}
\DeclareMathOperator{\R}{Re}
\DeclareMathOperator{\I}{Im}
\newcommand{\hol}{\text{hol}}
\newcommand{\BH}{\text{BH}}

\newcommand{\lit}{\ell}

\newcommand{\A}{I}
\newcommand{\B}{J}
\newcommand{\cN}{\mathcal{N}}

\newcommand{\K}{K}
\newcommand{\W}{W}
\newcommand{\Gamt}{\tilde{\Gamma}}

\newcommand{\Dbar}[2]{\bar{\mathcal{D}}_{\bar{#1}}\bar{#2}}
\newcommand{\Iprod}[2]{\langle {#1}, {#2} \rangle}
\newcommand{\abs}[1]{\lvert #1\rvert}
\newcommand{\db}{\mathbf{d}}
\newcommand{\xb}{\mathbf{x}}

\preprint{IFUM-946-FT, KUL-TF-09/19}

\title{Non-supersymmetric extremal multicenter black holes with superpotentials}

\author{Pietro Galli\\
Dipartimento di Fisica, Sezione Teorica,
Universit\`a degli Studi di Milano\\
Via Celoria, 16, 20133 Milano, Italy\\
\email{pietrogll@aol.com}}

\author{Jan Perz\\
Afdeling Theoretische Fysica, Katholieke Universiteit Leuven\\
Celestijnenlaan 200D bus 2415, 3001 Heverlee, Belgium\\
\email{Jan.Perz@fys.kuleuven.be}}

\abstract{Using the superpotential approach we generalize Denef's method of deriving and solving first-order equations describing multicenter extremal black holes in four-dimensional $\cN=2$ supergravity to allow non-supersymmetric solutions. We illustrate the general results with an explicit example of the $stu$ model.}


\begin{document}

\section{Introduction}

Most of insight we have gained into the origin of black hole entropy comes from the analysis of supersymmetric solutions in supergravity and string theory, an important class of which are multicenter black holes (the Majumdar--Papapetrou solutions \cite{Majumdar:1947eu,Papapetrou:1947ib} in the Einstein--Maxwell theory can be seen as their precursors). In four-dimensional $\cN=2$ supergravity coupled to vector multiplets the general stationary supersymmetric multi-black holes, also known as black hole composites, were obtained by Denef \cite{Denef:2000nb} (extending results of Behrndt, L\"ust and Sabra \cite{Behrndt:1997ny}; higher-curvature corrections were taken into account by Cardoso, de Wit, K\"appeli and Mohaupt \cite{LopesCardoso:2000qm}). To better understand black holes that are not supersymmetric, however, it is desirable to look for new solutions, in particular those that would still share certain features (such as extremality) with their supersymmetric counterparts, so that at least some of the tools developed for the latter could be applied to the former.

Recently two methods have been used to construct non-supersymmetric extremal multicenter solutions: Gaiotto, Li and Padi \cite{Gaiotto:2007ag}, following the earlier idea of Breitenlohner, Maison and Gibbons \cite{Breitenlohner:1987dg}, through dimensional reduction over the timelike Killing direction mapped a class of four-dimensional static multicenter black holes (which includes both supersymmetric and non-supersymmetric solutions) to geodesics on the scalar manifold, for the case when it is a symmetric coset space. The geodesics are then traced by the nilpotent generators of the coset algebra. A similar study, but carried out for maximal rather than $\cN=2$ supergravity coupled to a single vector multiplet, was later performed by Bossard and Nicolai \cite{Bossard:2009my}. The same type of dimensional reduction was also the main tool of the systematic study by Mohaupt and Waite \cite{Mohaupt:2009iq} of conditions under which static electric multicenter solutions in theories with Einstein--Maxwell-type Lagrangians in five spacetime dimensions can be expressed by harmonic functions.

Goldstein and Katmadas \cite{Goldstein:2008fq} in turn observed that one could break supersymmetry, but still satisfy the equations of motion of five-dimensional extremal supergravity solutions with a four-dimensional Gibbons--Hawking or Taub-NUT base space, by reversing the orientation of the base.\footnote{In fact the equations of motion will remain satisfied also after replacing the Euclidean four-dimensional hyper-K\"ahler base with a more general Ricci-flat space \cite{Bena:2009qv}.} By spacelike dimensional reduction these authors were able to obtain non-supersymmetric multicenter configurations also in four spacetime dimensions. Subsequently Bena et al.~\cite{Bena:2009ev,Bena:2009en} demonstrated examples of non-supersymmetric multicenter solutions with non-zero angular momentum and non-trivial constraints on the relative positions of the centers.

Meanwhile Gimon, Larsen and Sim\'on \cite{Gimon:2007mh,Gimon:2009gk}, motivated by the form of the ADM mass formula, provided an interpretation of a single-center extremal non-supersymmetric black hole in the $stu$ model as a threshold bound state (where the binding energy between the components vanishes) of four constituents, each of which is supersymmetric when considered individually.

Here, in the context of four-dimensional $\cN=2$ supergravity with cubic prepotentials, we present another way of obtaining extremal non-supersymmetric multicenter solutions, which directly generalizes Denef and Bates's original supersymmetric derivation \cite{Denef:2000nb,Bates:2003vx}, and which is an application of the superpotential approach, so far employed for single-center solutions \cite{Ceresole:2007wx,Andrianopoli:2007gt,Cardoso:2007ky,Perz:2008kh,Andrianopoli:2009je,Ceresole:2009iy,Bossard:2009we}. Figuratively speaking, this method consists in replacing the central charge in the equations governing the solution by a different, but typically very closely related quantity, known as the (fake) superpotential. To make the merger with Denef's formalism possible with minimal modification, we restrict ourselves to systems, which turn out to have constituents with mutually local charges.

Before explaining this technique in more detail in section \ref{sec:Core}, we will introduce the necessary concepts and notation in section \ref{sec:Formalism}. In section \ref{sec:stu-example} we (re-)derive simple examples of non-supersymmetric solutions in the $stu$ model: a single-center solution with non-vanishing central charge, first obtained by Tripathy and Trivedi \cite{Tripathy:2005qp}, and a multi-center solution, of the type conjectured by Kallosh, Sivanandam and Soroush \cite{Kallosh:2006ib}. We also mention how the BPS constituent interpretation fits into our framework. The final section \ref{sec:Conclusions} summarizes and discusses the results.

\section{Differential and special geometry}\label{sec:Formalism}

In this technical section we are going to briefly recall some basic concepts of special K\"ahler geometry \cite{Strominger:1990pd,Craps:1997gp}---the target space geometry of $\cN=2$ supergravity \cite{deWit:1984pk,deWit:1984px}---needed for finding single-center and multicenter charged extremal black hole solutions in four spacetime dimensions, following the formalism employed by Denef for the supersymmetric case. For a more exhaustive exposition we refer the reader to, for instance, \cite{Candelas:1990pi,Grimm:2004ua} and \cite{Pioline:2008zz}.

We can look at the four-dimensional theory from a higher-dimensional perspective.\footnote{Early papers on the subject of black hole composites, such as \cite{Denef:2000nb,Denef:2001xn,Bates:2003vx}, predominantly adopted type IIB interpretation; we choose type IIA, common in  more recent work, e.g.~\cite{deBoer:2008fk,VanHerck:2009ww}.} By compactifying six of the ten dimensions of type IIA string theory on a Calabi--Yau three-fold $X$ (or, equivalently, type IIB on the mirror of $X$) one finds \cite{Bodner:1990zm} an effective $\cN=2$ supergravity theory, whose bosonic sector is described by the action
\begin{equation}
\label{Ssugra4D}
\begin{split}
\Sa_\text{4D}=\frac{1}{16\pi}\int \Bigl(&R\Hodge 1 - 2 g_{a\bar{b}}(z,\bar{z})\,\de z^a\wedge\Hodge \de\bar{z}^{\bar{b}} \\ &+ \I \mathcal{N}_{\A\B}(z,\bar{z})\,\mathcal{F}^{\A}\wedge\Hodge\mathcal{F}^{\B} + \R\mathcal{N}_{\A\B}(z,\bar{z})\,\mathcal{F}^{\A} \wedge \mathcal{F}^{\B} \Bigr).
\end{split}
\end{equation}
In this action the field strengths are defined as $\mathcal{F}^{\A}=\de A^{\A}$ with the index ${\A}=\nobreak(0,a)$ labeling the abelian gauge fields $A^{\A}=(A^0,A^a)$ of the gravity multiplet and, respectively, the vector multiplets of the theory. The vector multiplets are enumerated by the Hodge number $h^{1,1}=\dim H^{1,1}(X)$. Each vector multiplet contains two neutral real scalars, combined into a complex scalar: $z^a=\mathcal{X}^a+\im\mathcal{Y}^a$. Hypermultiplets (and the tensor multiplet, which can be dualized to another hypermuliplet) do not play a role in our discussion, hence we have set them to zero.

The compactification manifold $X$ is characterized by its intersection numbers defined as:
\begin{equation}\label{intersectionNumber}
 D_{abc}=\int_X D_a\wedge D_b \wedge D_c \, ,
\end{equation}
where the set $\{D_a\}$ comprises a basis of $H^2(X)=H^{1,1}(X)$. Using this quantity, we introduce for any $\xi=\xi^aD_a$ the notation:
\begin{equation}
 \label{defX^3,X^2_a,X_ab}
\begin{split}
 \xi^3 &=\int_X \xi\wedge\xi\wedge\xi=D_{abc}\xi^a\xi^b\xi^c \,,\\
 \xi_a^2 &=\int_X D_a\wedge\xi\wedge\xi=D_{abc}\xi^b\xi^c \, ,\\
 \xi_{ab}&=\int_X D_a\wedge D_b\wedge\xi=D_{abc}\xi^c\, .
\end{split}
\end{equation}

The scalar manifold is special K\"ahler with the metric:
\begin{equation}
 \label{scalarMetric}
g_{a\bar{b}}=\frac{1}{4\mathcal{Y}^3}\int_X D_a\wedge\Hodge D_{\bar{b}}=-\frac{3}{2}\left( \frac{\mathcal{Y}_{a\bar{b}}}{\mathcal{Y}^3}-\frac{3}{2}\frac{\mathcal{Y}_a^2\mathcal{Y}^2_{\bar{b}}}{(\mathcal{Y}^3)^2}\right)=-\partial_{z^a}\partial_{\bar{z}^{\bar{b}}}\Bigl(\ln\frac{4}{3}\mathcal{Y}^3\Bigr).
\end{equation}
This equation shows that $g_{a\bar{b}}=\partial_a\bar{\partial}_{\bar{b}}\K$ is a K\"{a}hler metric with the K\"{a}hler potential $\K(z,\bar{z})=-\ln\frac{4}{3}\mathcal{Y}^3$. In fact both the K\"{a}hler potential and the vector couplings $\mathcal{N}_{\A\B}(z,\bar{z})$ appearing in \eqref{Ssugra4D} can be calulated from a single function, the holomorphic cubic prepotential, homogeneous of second degree in the projective coordinates $X^\A$ (such that $z^a=X^a/X^0$):
\begin{equation}
F=-\frac{1}{6}D_{abc}\frac{X^a X^b X^c}{X^0}=(X^0)^2 f(z)\,
\qquad f(z) = -\frac{1}{6}D_{abc} z^a z^b z^c\,.
\end{equation}

Many objects of relevance will be most naturally thought of as taking values in the even cohomology of the internal Calabi--Yau manifold $X$:
\begin{equation}
H^{2\ast}(X)=H^0(X)\oplus H^2(X)\oplus H^4(X)\oplus H^6(X)\,.
\end{equation}
The even cohomology has dimension $2h^{1,1}+2$ and each element $E\in H^{2\ast}(X)$ can be expanded as:
\begin{equation}
 E=E^0+E^aD_a+E_aD^a+E_0\de V\, .
\end{equation}
$\de V$ is the normalized volume form on $X$ and $\{D^a\}$ is a dual basis of $H^4(X)$ such that:
\begin{equation}
 \int_X D_{\A}\wedge D^{\B} = \delta_{\A}^{\B} \, .
\end{equation}

We will make use of the following antisymmetric topological intersection product of two polyforms belonging to $H^{2\ast}(X)$:
\begin{equation}
 \label{intersectionProd}
\langle E_1,E_2\rangle=\int_X E_1\wedge E_2^{\ast} \, ,
\end{equation}
where the action of the operator $^{\ast}$ on $E$ is simply a change of sign of the 2- and 6-form components. The intersection product in terms of components then reads:
\begin{equation} \label{IprodComponents}
 \langle E_1,E_2 \rangle=-E^0_1E_0^2+E_1^aE_a^2-E_a^1E_2^a+E_0^1E_2^0 \, .
\end{equation}

We define the \emph{period vector}, an object belonging to $H^{2\ast}(X)$  that entails the quantities introduced so far:
\begin{equation}
 \label{periodVector}
\Omega_{\hol}(z)=-1-z^a D_a-\frac{z^2_aD^a}{2} -\frac{z^3}{6}\de V \, .
\end{equation}
A normalized version of $\Omega_{\hol}$ satisfying $\langle \Omega,\bar{\Omega} \rangle=-\im$ is:
\begin{equation}
 \label{perioVectorNormalized}
\Omega(z,\bar{z})=\ee^{\K/2}\Omega_{\hol}= \sqrt{\frac{3}{4\mathcal{Y}^3}}\;\Omega_{\hol}\, .
\end{equation}

The period vector $\Omega$ transforms under K\"{a}hler transformations with K\"{a}hler weight $(1,-1)$ and we want its derivative to transform in the same way. To achieve this we define its covariant derivatives as:
\begin{equation}
 \label{covariantDerivativeOmega}
\begin{split}
\mathcal{D}_a\Omega&=\partial_a \Omega + \tfrac{1}{2} \partial_a \K\,\Omega \, ,\\
\bar{\mathcal{D}}_{\bar{a}}\Omega &=
\bar{\partial}_{\bar{a}} \Omega - \tfrac{1}{2} \bar{\partial}_{\bar{a}} \K\,\Omega=0 \, .
\end{split}
\end{equation}
The second relation expresses the covariant holomorphicity of the normalized period vector with respect to the K\"{a}hler connection.

Using the normalized period vector one can associate a new quantity, which we decide to call \emph{fake central charge function}, to every element  $E\in H^{2\ast}(X)$:
\begin{equation}
 \label{Z(E)def}
Z(E)=\Iprod{E}{\Omega}= \sqrt{\frac{3}{4\mathcal{Y}^3}}\,\left(\frac{E^0z^3}{6}-\frac{E^a z^2_a}{2}+E_az^a-E_0\right).
\end{equation}
The name ``fake central charge function'' is given because, when $E$ encodes the electromagnetic charges carried by the vector fields, this object becomes the central charge function which, at spatial infinity, is the true central charge of the relevant four dimensional supersymmetry algebra.

With the definitions above, the set $\{\Omega,\mathcal{D}_a\Omega,\bar{\mathcal{D}}_{\bar{a}}\bar{\Omega},\bar{\Omega}\}$ constitutes an alternative basis of $H^{2\ast}(X)$. In fact one can prove the validity of the following equalities:
\begin{equation}
 \begin{split}
  \langle&\Omega,\bar{\Omega} \rangle = -\im\, ,\\
  \langle&\mathcal{D}_a\Omega,\Dbar{b}{\Omega} \rangle = \im g_{a\bar{b}}\, ,\\
  \langle&\mathcal{D}\Omega,\Omega \rangle = 0\, .
 \end{split}
\end{equation}
In this new basis a constant real element $E\in H^{2\ast}(X)$ can be expanded as:
\begin{equation}
 \label{EonOmegaBas}
\begin{split}
E &= \im\bar{Z}(E)\Omega - \im g^{\bar{a}b}\Dbar{a}{Z}(E)\mathcal{D}_b\Omega + \im g^{a\bar{b}}\mathcal{D}_a Z(E)\Dbar{b}{\Omega} - \im Z(E)\bar{\Omega} \\
&=-2\I\bigl[\bar{Z}(E)\Omega-g^{\bar{a}b}\Dbar{a}{Z}(E)\mathcal{D}_b\Omega\bigr]\, .
\end{split}
\end{equation}

Let us finally introduce the operator $\diam$ acting on the basis elements in the following way:
\begin{equation}
 \diam\Omega=-\im\Omega\,,\qquad
 \diam\bar\Omega=\im\bar\Omega\,,\qquad
 \diam\mathcal{D}_a\Omega=\im\mathcal{D}_a\Omega\,,\qquad
 \diam\Dbar{a}{\Omega}=-\im\Dbar{a}{\Omega}\,.
\end{equation}
Using this new operator one can define a positive non-degenerate norm on $H^{2\ast}(X,\mathbb{R})$ as:
\begin{equation}
 \label{NormIProd}
\lvert E \rvert^2 = \Iprod{E}{\diam E} \, .
\end{equation}

\section{Extremal black holes with a superpotential}\label{sec:Core}

\subsection{Single-center black holes}

Before generalizing to multicenter black holes let us consider the case in which all the electromagnetic charges are carried by a single center and let us assume spherical symmetry. All the quantities (scalars as well) depend thus only on the radial coordinate $r$ or equivalently on $\tau=\frac{1}{\lvert r-r_\mathrm{h}\rvert}$. The ansatz for a static metric is:
\begin{equation}
 \label{metricSC}
\de s^2=-\ee^{2U}\de t^2 + \ee^{-2U}\delta_{ij}\de x^i\de x^j\,,
\end{equation}
with $U=U(r)$ called warp factor. Requiring asymptotically flat metric imposes the constraint $U_{r\to\infty}=U_{\tau\to0}\to 0$.

The electromagnetic field strength $\mathcal{F}$ consistent with symmetries is:
\begin{equation} \label{curlyF}
 \mathcal{F}=\mathcal{F}_{\text{m}}+\mathcal{F}_{\text{e}}=\sin\theta\,\de\theta\wedge\de\varphi \otimes \Gamma + \ee^{2U}\de t\wedge\de\tau \otimes \diam\Gamma\, ,
\end{equation}
with the components of the polyform
\begin{equation}
 \Gamma=\Gamma(Q)=p^0+p^aD_a+q_aD^a+q_0\de V
\end{equation}
encoding the charges carried by the black hole (or, in the geometrical interpretation, numbers of D-branes wrapping even cycles of the compactification manifold $X$), which we could alternatively arrange in a symplectic vector $Q=(p^\A,q_\B)$.

Under these assumptions, the total action \eqref{Ssugra4D} in terms of $\tau$ and per unit time can be recast in the form \cite{Ferrara:1997tw}:
 \begin{equation}
 \label{Seff}
 \Sa_{\text{eff}} = -\frac{1}{2} \int^{\infty}_{0}\de\tau\,\bigl(\dot{U}^2 + g_{a\bar{b}} \dot{z}^a \dot{\bar{z}}^{\bar{b}} + \ee^{2U}V_{\BH}\bigr) - (\ee^U \abs{Z})_{\tau=\infty}\, .
\end{equation}
Here we have neglected the boundary term proportional to $\dot{U}$ and used the shorthand notation $Z=Z(\Gamma)$. The dot indicates differentiation with respect to $\tau$ and the effective black hole potential is given by:
\begin{equation}
\label{BHpotZ}
V_{\BH}=\tfrac{1}{2}\Iprod{\Gamma}{\diam\Gamma}=\abs{Z}^2+4g^{a\bar{b}}\,\partial_a\abs{Z}\bar{\partial}_{\bar{b}}\abs{Z}\, .
\end{equation}

The black hole potential \eqref{BHpotZ} is a quadratic polynomial in the charges and can be expressed as $V_{\BH}=Q^{\text{T}}\mathcal{M}Q$ with a certain  matrix $\mathcal{M}$. We have the freedom to perform transformations on the charge vector  $Q \rightarrow SQ$ without changing the value of $V_{\BH}$. This freedom lies in the possibility to choose the symplectic matrix $S$ among all those that satisfy \cite{Ceresole:2007wx}:
\begin{equation}
\label{VBH_S}
 V_{\BH}=Q^{\text{T}}\mathcal{M}Q=Q^{\text{T}}S^{\text{T}}\mathcal{M}SQ\quad\Rightarrow\quad S^{\text{T}}\mathcal{M}S=\mathcal{M}\, .
\end{equation}

The sum of squares \eqref{BHpotZ} is therefore not unique and one can more generally consider the effective action
\begin{equation}
\label{Seff_W}
 \Sa_{\text{eff}} = -\frac{1}{2} \int^{\infty}_{0}\de\tau\, \left(\dot{U}^2 + g_{a\bar{b}} \dot{z}^a \dot{\bar{z}}^{\bar{b}} + \ee^{2U}(\W^2+4g^{a\bar{b}}\partial_a\W\bar{\partial}_{\bar{b}}\W)\right) - (\ee^U\W)_{\tau=\infty}\, ,
\end{equation}
with $\W$, usually called the fake superpotential, not necessarily equal to $\abs{Z}$.

Varying the action we obtain the following first order equations, by construction equivalent to the second order equations of motion:
\begin{align}
 \label{1_ord_UW}
\dot{U}&=-\ee^U\W\, ,\\
 \label{1_order_zW}
 \dot{z}^a&=-2\ee^U g^{a\bar{b}}\bar{\partial}_{\bar{b}}\W\, .
\end{align}
When $\W$ is equal to $\abs{Z(\Gamma)}$ \eqref{1_ord_UW} and \eqref{1_order_zW} describe a supersymmetric attractor flow \cite{Ferrara:1995ih,Ferrara:1997tw}. (The name ``attractor'' stems from the fact that the flow has a fixed point determined by the charges, which is reached by the scalars as they approach the event horizon, i.e.~when $\tau\to\infty$.) When $\W\neq\abs{Z(\Gamma)}$ the flow is non-supersymmetric.

The form \eqref{1_ord_UW} and \eqref{1_order_zW} of the attractor equations emphasizes the gradient nature of the flow, but to be able to integrate them directly, another form is more suitable. In the supersymmetric case it follows from the rewriting of the action in yet another way \cite{Denef:2000nb} (but still as a sum of squares):
\begin{equation}
\label{Seff_BPS}
 \Sa_{\text{eff}}=-\frac{1}{4}\int_0^\infty\de\tau\,\ee^{2U}\Bigl|2\I\bigl[(\partial_\tau+\im\mathcal{Q}_\tau + \im\dot{\alpha})(\ee^{-U}\ee^{-\im\alpha}\Omega)\bigr]+\Gamma\Bigr|^2
- (\ee^U\abs{Z})_{\tau=0} \, ,
\end{equation}
where $\mathcal{Q}_\tau=\I(\partial_a\K\dot{z}^a)$ and $\alpha=\arg Z(\Gamma)$.

Based on the similarity between supersymmetric and non-supersymmetric equations, we generalize this expression by replacing $\Gamma$ with a different real element of the even cohomology of $X$, say $\Gamt$. To retain the same form of the expansion \eqref{EonOmegaBas} as that employed in supersymmetric solutions,
\begin{equation} \label{GamT_onOmegaBasis}
\Gamt = \im\bar{Z}(\Gamt)\Omega - \im g^{\bar{a}b}\Dbar{a}{Z}(\Gamt)\mathcal{D}_b\Omega + \im g^{a\bar{b}}\mathcal{D}_a Z(\Gamt)\Dbar{b}{\Omega} - \im Z(\Gamt)\bar{\Omega}\,,
\end{equation}
we limit our analysis to those $\Gamt$ that have constant real components in the basis $\{D_{\A},D^{\B}\}$. We can arrange these them in a symplectic vector $\tilde{Q}=(\tilde{p}^{\A},\tilde{q}_{\B})$, where $\tilde{Q}=SQ$, so that $\Gamt=\Gamma(\tilde{Q})$. Consequently, the matrix $S$ is restricted to be real and constant.\footnote{Already in \cite{Ceresole:2007wx} it was argued that only a constant matrix $S$ would allow the rewriting of $V_\BH$ in \eqref{VBH_S} as a sum of squares in terms of a superpotential obtained from the central charge by acting with $S$ on $Q$. In our formalism, if $S$ were moduli-dependent, the coefficient of $\mathcal{D}_b\Omega$ (and $\bar{\mathcal{D}}_{\bar{b}}\bar\Omega$) in the expansion \eqref{GamT_onOmegaBasis} would have an additional term (namely $-\langle\partial_a SQ,\Omega\rangle$), and expressing the effective action in a manner analogous to \eqref{Seff_BPS} would not be straightforwardly possible.}

With $\Gamt$ chosen in this way, we can identify the superpotential with the fake central charge function defined in \eqref{Z(E)def}, when evaluated for $\Gamt$ as its argument:
\begin{equation} \label{W=Ztilde}
 \W=\abs{Z(\Gamt)}\,.
\end{equation}
It follows that
\begin{equation}
 \abs{\Gamt}^2=\W^2+4g^{a\bar{b}}\partial_a\W\bar{\partial}_{\bar{b}}\W\,.
\end{equation}
Denoting $\tilde{\alpha}=\arg Z(\Gamt)$ we can write the effective action \eqref{Seff_W} as
\begin{equation}
\label{Seff_nBPS}
 \Sa_{\text{eff}}=-\frac{1}{4}\int^\infty_0\de\tau\,\ee^{2U}\left|2\I\bigl[(\partial_\tau+\im\mathcal{Q}_\tau+\im\dot{\tilde{\alpha}})(\ee^{-U}\ee^{-\im\tilde{\alpha}}\Omega)\bigr]+\Gamt\right|^2
- (\ee^U\W)_{\tau=0} \, ,
\end{equation}
and the attractor equations become, in complete analogy with Denef's original treatment of the supersymmetric case:
\begin{equation}
 \label{Im=Gamt}
2\partial{_\tau}\I(\ee^{-U}\ee^{-\im\tilde{\alpha}}\Omega)=-\Gamt\,.
\end{equation}

The form \eqref{Im=Gamt} of the attractor equations is suitable for direct integration  and gives:
\begin{equation}
\label{Im=Htilde}
  2\ee^{-U}\I(\ee^{-\im\tilde{\alpha}}\Omega)=-\tilde{H}(\tau)\, ,
\end{equation}
with
\begin{equation} \label{Htilde}
\tilde{H}(\tau)=\Gamt\tau - 2\I(\ee^{-\im\tilde{\alpha}}\Omega)_{\tau=0}\, .
\end{equation}
The explicit solution for the scalars is \cite{Bates:2003vx}:
\begin{equation}
 \label{z_nBPS}
z^a(\tilde{H})=\frac{\tilde{H}^a-\im\,\de_{\tilde{H}_a}\Sigma(\tilde{H})}{\tilde{H}^0+\im\,\de_{\tilde{H}_0}\Sigma(\tilde{H})}\, ,
\end{equation}
and
\begin{equation}
\ee^{-2U(\tilde{H})} = \lvert Z(\tilde{H})\rvert^2\Bigr|_{z=z(\tilde{H})} =\W^2(\tilde{H})\Bigr|_{z=z(\tilde{H})} = \Sigma(\tilde{H})\,,
\end{equation}
where the entropy function $\Sigma(\tilde{H})$ can be obtained, as in the supersymmetric case, from the entropy of the black hole
\begin{equation}
 \mathcal{S}_{\BH}=\pi\Sigma(\Gamt)=\pi\W^2(\Gamt)
\end{equation}
by replacing the charges with harmonic functions.

The question we are left with thus concerns the conditions allowing the existence of the constant matrix $S$. They may be met by truncating the theory to a suitable subset of the scalar fields. In particular for the $stu$ model it has been shown  to mean setting to zero the axion fields $\R z^a$ and considering magnetic or electric configurations \cite{Ceresole:2007wx}. This assumption is the same for the $t^3$ and $st^2$ models\footnote{It suffices to compare the BPS and non-BPS attractor solutions of the $t^3$ or $st^2$ model \cite{Ceresole:2007wx,Bellucci:2007zi}. As for the $stu$ model, once we impose $\R z^a=\mathcal{X}^a=0$ and consider the magnetic or electric configuration, the solutions differ only by a switch of sign of the charges---one that would be effected by the matrix $S$ in our treatment.} and all models with cubic prepotentials. In this setting $S$ turns out to be diagonal and acts on the charge vector without changing its electric or magnetic character.

In what follows we will assume the scalars to be purely imaginary and the charge configuration to be either $(p^0,0,0,q_a)$ or $(0,p^a,q_0,0)$.

\subsection{Multicenter black holes} \label{subsec:Multicenter non-BPS}

For multicenter configurations the spherical symmetry assumption of the previous derivation is no longer valid and we have to consider more general, stationary spacetimes, by including in the metric an extra one-form $\omega=\omega_i\de x^i$:
 \begin{equation}
 \label{metricMltc}
 \de s^2=-\ee^{2U}(\de t + \omega_i\de x^i)^2 + \ee^{-2U}\delta_{ij}\de x^i\de x^j\,
\end{equation}
and taking $U$ and $\omega_i$ to be arbitrary functions of position $\xb$. We require asymptotic flatness by imposing $U,\omega\rightarrow 0$ when $\tau\rightarrow 0$.

Although the idea for obtaining the attractor flow equations remains the same (namely the rewriting of the Lagrangian as a sum of squares), the formalism becomes more involved. Following with some alterations  reference \cite{Denef:2000nb}, we adopt the boldface notation for three-dimensional quantities. The 3D Hodge dual with respect to the flat metric $\delta_{ij}$ will be denoted by $\boldsymbol{\Hodge_0}$ and for convenience we define $w=\ee^{2U}\omega$. We also need to introduce the following scalar product of spatial 2-forms $\boldsymbol{\mathcal{F}}$ and $\boldsymbol{\mathcal{G}}$:
\begin{equation}
 \label{Prod2forms}
 (\boldsymbol{\mathcal{F}},\boldsymbol{\mathcal{G}}) = \frac{\ee^{2U}}{1-w^2} \int_X \boldsymbol{\mathcal{F}}\wedge\big[
 \boldsymbol{\Hodge_0}(\diam \boldsymbol{\mathcal{G}}^{\ast}) -
 \boldsymbol{\Hodge_0}(w\wedge\diam\boldsymbol{\mathcal{G}}^{\ast})\,w
 +\boldsymbol{\Hodge_0}(w\wedge\boldsymbol{\Hodge_0}\boldsymbol{\mathcal{G}}^{\ast})\big]\, ,
\end{equation}
where $^{\ast}$ is an operator acting on the elements of the even cohomology of $X$ as defined below formula \eqref{intersectionProd}. The product just introduced is  commutative and we can assume it to be positive definite taking $w$ small enough.

With this notation the effective action reads (dropping the total derivative $\Delta U$):
\begin{equation}
\label{S4Dmltc}
\begin{split}
 \Sa_{\text{4D eff}}=-\frac{1}{16\pi}\int\de t \int_{\mathbb{R}^3}\Big[&2\db U\wedge\boldsymbol{\Hodge}_\mathbf{0}\db U - \tfrac{1}{2}\ee^{4U}\db\omega\wedge\boldsymbol{\Hodge}_\mathbf{0}\db\omega
\\&+2g_{a\bar{b}}\,\db z^a\wedge\boldsymbol{\Hodge}_\mathbf{0}\db \bar{z}^{\bar{b}}+(\boldsymbol{\mathcal{F}},\boldsymbol{\mathcal{F}})\Big].
\end{split}
\end{equation}

Generalizing Denef's derivation in the way we did for single-center black holes, we introduce the electromagnetic field strength corresponding to the modified charges $\Gamt$:
\begin{equation}
 \label{Ftilde}
 \frac{1}{4\pi}\int\tilde{\boldsymbol{\mathcal{F}}}=\Gamt\, .
\end{equation}
Consequently we define
\begin{equation}
 \tilde{\boldsymbol{\mathcal{G}}}=\tilde{\boldsymbol{\mathcal{F}}}-2\I\boldsymbol{\Hodge}_\mathbf{0}\mathbf{D}(\ee^{-U}\ee^{-\im \tilde{\alpha}} \Omega )+2\R\mathbf{D}(\ee^{U}\ee^{-\im\tilde{\alpha}}\Omega\,\omega)
\end{equation}
and write the Lagrangian of \eqref{S4Dmltc} in the form\footnote{We assume that the constraints \cite{VanProeyen:2007pe} resulting from the components of Einstein's equation not reproduced by this Lagrangian will remain satisfied as in the supersymmetric case.}
\begin{equation}
\begin{split}
 \mathcal{L}={}&(\tilde{\boldsymbol{\mathcal{G}}},\tilde{\boldsymbol{\mathcal{G}}})\,-\,4\,(\mathbf{Q}+\db\tilde{\alpha}+\tfrac{1}{2}\ee^{2U}\boldsymbol{\Hodge}_\mathbf{0}\db\omega)\wedge\I\langle\tilde{\boldsymbol{\mathcal{G}}},\ee^U \ee^{-\im\tilde{\alpha}}\Omega\rangle\\
&+\db\,[2w\wedge(\mathbf{Q}+\db\tilde{\alpha})+4\R\langle\tilde{\boldsymbol{\mathcal{F}}},\ee^U\ee^{-\im\tilde{\alpha}}\Omega\rangle]\, .
\end{split}
\end{equation}
with
\begin{align}
 \label{Dmltc}
  \mathbf{D}&=\db+\im(\mathbf{Q}+\db\alpha+\tfrac{1}{2}\ee^{2U}\boldsymbol{\Hodge_0}\db\omega)\, ,
 \\ \label{QmltcAlpha}
  \mathbf{Q}&=\I\,(\partial_a\K\db z^a)\, .
\end{align}
 Imposing the first order equations
\begin{align}
 \label{G=0nBPS}
 \tilde{\boldsymbol{\mathcal{G}}}&=0\, ,\\
 \label{Q+da+1/2e=0nBPS}
 \mathbf{Q}+\db\tilde{\alpha}+\tfrac{1}{2}\ee^{2U}\boldsymbol{\Hodge_0}\db\omega&=0
\end{align}
solves the equations of motion. From \eqref{Q+da+1/2e=0nBPS} it follows that $\mathbf{D}=\db$ and then, as by definition and our assumption $\db\tilde{\boldsymbol{\mathcal{F}}}=0$, differentiating  \eqref{G=0nBPS} leads to
\begin{equation}
 \label{laplacian=0_nBPS}
2\db\boldsymbol{\Hodge_0}\db\I(\ee^{-U}\ee^{-\im\tilde{\alpha}}\Omega)=0\, .
\end{equation}
This is a Laplacian equation which integrated gives (cf.~\eqref{Im=Htilde}):
 \begin{equation}
  \label{Im=Hmltc_nBPS}
 2\ee^{-U}\I(\ee^{-\im\tilde{\alpha}}\Omega)=-\tilde{H}\, ,
 \end{equation}
where $\tilde{H}$ is a generic $H^{2\ast}(X)$-valued harmonic function. Since we are looking for non-BPS multicenter configurations considering $N$ sources at position $\xb_n$, it seems reasonable to take as $\tilde{H}(\xb)$ a natural generalization of \eqref{Htilde}, namely:
 \begin{equation}
  \label{Hmltc_nBPS}
 \tilde{H}(\xb)=\sum_{n=1}^N\Gamt_n\tau_n-2\I(\ee^{-\im\tilde{\alpha}}\Omega)_{\tau=0}\,,
 \end{equation}
with $\tau_n=\abs{\xb-\xb_n}^{-1}$ and $\Gamt_n=\Gamma(S_n Q_n)$, $S_n$ being constant matrices.

To be able to speak of black hole composites, certain conditions need to be satisfied:
\begin{itemize}
\item A single-center non-BPS black hole of total charge $Q$ and its corresponding attractor flow have to exist and be well defined  (i.e.\ they have to be describable with the above procedure).
\item For each center of charge $Q_n$ a single-center attractor flow has to exist as well.
\item  The charges must obey the constraints:
\begin{align}
 Q&=\sum_{n=1}^N Q_n\, ,\label{chargeConstraint}\\
 \Gamt&=\Gamma(SQ)=\sum_{n=1}^N\Gamma(S_n Q_n)=\sum_{n=1}^N\Gamt_n\, .\label{SConstraint}
\end{align}
\end{itemize}

In addition we need to take into account a particular feature of the central charge $Z$, stemming from our assumptions regarding the charges: taking $Q=(p^0,0,0,q_a)$ or $Q=(0,p^a,q_0,0)$ and imposing $\R z^a=0$ reveals the central charge to have a constant phase. For instance, with $Q$ electric the corresponding central charge reads
\begin{equation}
 Z(\Gamma)= \sqrt{\frac{3}{4\mathcal{Y}^3}}\,\left(\frac{p^0 z^3}{6}+ q_a z^a\right)
\end{equation}
and with $z^a=\im\mathcal{Y}^a$ (where $\mathcal{Y}^a\in\mathbb{R}$) it holds that $\ee^{\im\alpha}=\frac{Z}{\lvert Z \rvert}=\im$.

Note that this is true also for  $\tilde{Z}=Z(\Gamt)$ whenever the difference between $Q$ and $\tilde{Q}$ amounts to constant factors multiplying their components (as is the case when $S$ is a constant diagonal matrix).
Then, as a direct consequence of the constancy of $\tilde{\alpha}$, it follows in our treatment that $\db\tilde{\alpha}=0$ and \eqref{Q+da+1/2e=0nBPS} in particular becomes:
\begin{equation}
 \label{Q=dw_nBPS}
 \mathbf{Q}=-\frac{1}{2}\ee^{2U}\boldsymbol{\Hodge_0}\db\omega\, .
\end{equation}

\subsubsection*{A different form of flow equations}

Let us bring the attractor equations \eqref{Im=Hmltc_nBPS} to a form more closely resembling the first order flow equations for the scalars and the warp factor (\ref{1_ord_UW}, \ref{1_order_zW}). In view of this we define:
\begin{equation}
 \tilde{\boldsymbol{\xi}}= \langle\db \tilde{H},\Omega\rangle=\sum_{n=1}^N Z(\tilde{\Gamma}_n)\,\db\tau_n = \sum_{n=1}^N\ee^{\im\tilde{\alpha}_n}\W_n\,\db\tau_n\, .
\end{equation}
Let us differentiate \eqref{Im=Hmltc_nBPS} to obtain:
\begin{equation}\label{dd}
\begin{split}
 \db\tilde{H}&=2\I\bigl[(\db U\Omega-\db\Omega)\ee^{-U}\ee^{-\im\tilde{\alpha}}\bigr]\\
&=2\I\bigl[(\db U\Omega-\mathcal{D}_a\Omega\,\db z^a + \im\mathbf{Q}\Omega)\ee^{-U}\ee^{-\im\tilde{\alpha}}\bigr]\,.
\end{split}
\end{equation}
Taking now the intersection product of \eqref{dd} with $\Omega$ yields:
\begin{equation}
 -\tilde{\boldsymbol{\xi}}=(\db U-\im\mathbf{Q})\ee^{-U}\ee^{\im\tilde{\alpha}}\,,
\end{equation}
and then:
\begin{align}
 \label{Q=I_nBPS}
\mathbf{Q}&=\ee^U\I(\ee^{-\im\tilde{\alpha}}\tilde{\boldsymbol{\xi}})\, , \\
\label{1order_Umltc_nBPS}
 \db U&=-\ee^U\R(\ee^{-\im\tilde{\alpha}}\tilde{\boldsymbol{\xi}})\, .
\end{align}
Similarly taking the intersection product of \eqref{dd} with $\bar{\mathcal{D}}_{\bar{a}}\bar{\Omega}$ gives:
\begin{equation}\label{z_mltc_nBPS}
 \db z^a=-\ee^{U}g^{a\bar{b}}\ee^{\im\tilde{\alpha}}\bar{\mathcal{D}}_{\bar{b}}\bar{\tilde{\boldsymbol{\xi}}}\,.
\end{equation}
Equations \eqref{1order_Umltc_nBPS}--\eqref{z_mltc_nBPS} are the multicenter version of \eqref{1_ord_UW}--\eqref{1_order_zW}. Recalling our assumptions and in particular using $\R z^a=0$ we have:
\begin{equation} \label{Q=0}
 \mathbf{Q}=\I(\partial_a\K\db z^a)=-\frac{\im}{2}(\partial_a\K\db z^a-\bar{\partial}_{\bar{a}}\K\db \bar{z}^{\bar{a}})=-\frac{\im}{2}(\partial_a\K\db z^a-\partial_a\K\db z^a)=0\,.
\end{equation}
Hence, with $\tilde\alpha_n = \arg Z(\Gamt_n)$, \eqref{Q=I_nBPS} becomes:
\begin{equation}\label{pp}
 0=\I(\ee^{-\im\tilde{\alpha}}\tilde{\boldsymbol{\xi}})=\sum_{n=1}^N\I\bigl(\ee^{-\im(\tilde{\alpha}-\tilde{\alpha}_n)}\bigr)\W_n\,\db\tau_n\,,
\end{equation}
that is $\tilde{\alpha}=\tilde{\alpha}_n\pmod\pi$ for all $n$.

\subsubsection*{Angular momentum and positions of the centers}

It is worth pointing out that equation \eqref{Q=0} applied to \eqref{Q=dw_nBPS} yields $\boldsymbol{\Hodge_0}\db\omega=0$, implying that the angular momentum $\mathbf{J}$, read off from the metric components as (see e.g.~\cite{Misner:1974qy}, ch.~19)
\begin{equation}
\omega_i = 2\epsilon_{ijk}J^j\frac{x^k}{r^3} + O(1/r^3) \quad \text{for} \quad r\to\infty\,,
\end{equation}
has to vanish and so the metric is in fact static. This is a remarkable difference with respect to the supersymmetric case, where, instead, the one-form $\omega$ enclosing the off-diagonal element of the metric is determined by solving equation \cite{Denef:2000nb}
\begin{equation}
 \boldsymbol{\Hodge_0}\db\omega=\Iprod{\db H}{H}\, .
\end{equation}

According to equation \eqref{pp}, the ``tilded'' central charges $\tilde{Z}=Z(\Gamt)$ and $\tilde{Z}_n=Z(\Gamt_n)$ have to be aligned either parallel or antiparallel. These are conditions analogous to those defining marginal or antimarginal stability in the BPS case. If we want to use the same terminology, this means that multicenter non-BPS systems described in this paper are marginally (or antimarginally) stable and can decompose into their constituents everywhere in moduli space. In the supersymmetric sector such a decay is for generic charge configurations possible only on a particular surface of the scalar manifold (the wall of marginal stability).

The relative positions of the sources in space are governed by the analogue of equation (7.23) in \cite{Denef:2000nb}:
\begin{equation}\label{constr_pos_nBPS}
 \sum_{n=1}^N\frac{\langle\Gamt_m,\Gamt_n\rangle}{\rvert \xb_m-\xb_n\lvert}=2\I\bigl[\ee^{-\im\tilde{\alpha}}Z(\Gamt_m)\bigr]_{\tau=0}\, .
\end{equation}
In the supersymmetric sector one finds $N-1$ constraints, which may even determine a nontrivial topology of the solution space \cite{deBoer:2008zn}. Here instead, since $\tilde{\alpha}=\tilde{\alpha}_m\pmod\pi$ for all $m$ implies $\I[\ee^{-\im\tilde{\alpha}}Z(\tilde{\Gamma}_m)]=0$ and then
\begin{equation}
 \sum_{n=1}^N\frac{\langle\Gamt_m,\Gamt_n\rangle}{\rvert \xb_m-\xb_n\lvert}=0 \, ,
\end{equation}
equation \eqref{constr_pos_nBPS} gives:
\begin{equation} \label{localCharge}
 \langle\tilde{\Gamma}_m,\tilde{\Gamma}_n\rangle=0\quad\forall\;m,n\,.
\end{equation}
This result in our context directly holds also for the charges $\Gamma_n$, stating that they have to be mutually local with respect to the product \eqref{intersectionProd}. Indeed, to satisfy the condition of constancy of $S$, we chose to work with electric or magnetic configurations, which lead to mutually local electric or magnetic constituents. As a consequence, there are no constraints on the positions and the centers are free.

\section{Non-BPS composites in the $stu$ model}\label{sec:stu-example}

In this section we are going to apply the general procedure described above to the particular case of the $stu$ model, as a concrete example. In this extensively studied model (see eg.~\cite{Bellucci:2008sv} and references therein), arising in type IIA compactification on a $T^2\times T^2 \times T^2$, the scalar manifold is the homogeneous symmetric space $\left(\frac{\mathrm{SU}(1,1)}{\mathrm{U}(1)}\right)^3$ parameterized by the complex moduli $z^1\equiv s$, $z^2\equiv t$ and $z^3\equiv u$ (corresponding to the complexified volumes of the tori). The prepotential reads:
\begin{equation}\label{Fstu}
 f=stu\,.
\end{equation}

The Bekenstein-Hawking entropy of a $stu$ black hole with charge\footnote{To match conventions used in some $stu$ literature, we have introduced the vector $Q_{\lit}$, differing from $Q$ by a sign reversal in the electric charges: $q^{\lit}_{a}=-q_{a}$.}
$Q_{\lit}=(p_{\lit}^{\A},q^{\lit}_{\B})$ is related through
\begin{equation}
 \mathcal{S}=\frac{A_{\text{h}}}{4}=\pi V_{\BH}\Bigr|_{\partial V_{\BH}=0}=\pi\sqrt{\abs{\mathcal{I}_4(Q_{\lit})}}
\end{equation}
to the unique invariant $\mathcal{I}_4$ of the tri-fundamental representation $(\mathbf{2,2,2})$ of the duality group $(\mathrm{SL}(2,\mathbb{Z}))^3$. Explicitly this invariant has the form:
\begin{equation}\label{I_4invariant}
\mathcal{I}_4(Q_{\lit})=-(p^{\A}_{\lit}q_{\A}^{\lit})^2+4\sum_{a<b}p^a_{\lit}q_a^{\lit}p^b_{\lit}q_b^{\lit}-4p^0_{\lit}q_1^{\lit}q_2^{\lit}q_3^{\lit}+4q_0^{\lit}p^1_{\lit}p^2_{\lit}p^3_{\lit}\,.
\end{equation}
Non-BPS black holes with $Z\neq0$ satisfy $\mathcal{I}_4(Q_{\lit})<0$.

Once we have chosen to deal with an electric charge configuration,\footnote{For a more generic non-BPS charge configuration one can apply an $\mathrm{SL}(2,\mathbb{Z})$ duality transformation, see e.g.~\cite{Gimon:2007mh}.} it follows that $\tilde{Q}=SQ=(-p^0,0,0,q_a)=(-p_{\lit}^0,0,0,-q_a^{\lit})$ and we can derive the non-BPS scalar solutions for single-center and multicenter black holes using the equations of our formulation.

In the single-center case we have to use the harmonic function (written here as a symplectic vector)
\begin{equation}
\tilde{H}=\begin{pmatrix} \tilde{p}^0 \\ \tilde{p}^a \\ \tilde{q}_0 \\ \tilde{q}_a \end{pmatrix} \tau + \tilde{h}_{\infty}=
\begin{pmatrix} -p_{\lit}^0 \\ 0 \\ 0 \\ -q^{\lit}_a \end{pmatrix} \tau + h^{\lit}_{\infty}\,,
\end{equation}
where with $h_\infty$ we have indicated the constant vector which at the end determines the value of the scalars at infinity. From \eqref{z_nBPS}, using $\Sigma^2(Q)=\mathcal{I}_4(Q_{\lit})$, we obtain the scalar solutions:
\begin{equation}
z^a(\tau)=\frac{-\im\,\de_{\tilde{H}_1}\Sigma(\tilde{H})}{\tilde{H}^0}
=\frac{-\im\,\de_{\tilde{H}_1}\sqrt{4\tilde{H}^0\tilde{H}_1\tilde{H}_2\tilde{H}_3}}{\tilde{H}^0}
=\frac{-\im\,\de_{H^{\lit}_1}\sqrt{4H_{\lit}^0H^{\lit}_1H^{\lit}_2H^{\lit}_3}}{H_{\lit}^0}
\end{equation}
and then
\begin{equation} \label{zSolelectric}
z^1(\tau)=-\im\,\sqrt{\frac{H_2^{\lit}H_3^{\lit}}{H^0_{\lit}H_1^{\lit}}}\,,\qquad
z^2(\tau)=-\im\,\sqrt{\frac{H_1^{\lit}H_3^{\lit}}{H^0_{\lit}H_2^{\lit}}}\,,\qquad
z^3(\tau)=-\im\,\sqrt{\frac{H_1^{\lit}H_2^{\lit}}{H^0_{\lit}H_3^{\lit}}}\,.
\end{equation}
These expressions correctly reproduce the results known from the existing literature \cite{Tripathy:2005qp,Kallosh:2006ib}.

The multicenter case is slightly more complicated. As we mentioned, a composite with $N$ centers of charge $Q_n$ at positions $\xb_n$ has to satisfy the constraints \eqref{chargeConstraint} and \eqref{localCharge}. In addition, at each $\xb_n$, there has to exist a single-center black hole described in terms of a harmonic function $\tilde{H}(\tau)$ of the form \eqref{Htilde}. The charge cofiguration at each of the $N$ centers needs to be either electric or magnetic, as these are the only configurations that allow non-BPS attractors describable with our procedure. However, since for both these configurations the matrix $S$ is diagonal, the constraints \eqref{chargeConstraint} are satisfied only if all $Q_n$ are of the same kind as $Q$. The composite is then constituted by $N$ single-center black holes with charge $Q_n=(p_n^0,0,0,q^n_a)=(p_{n\,\lit}^0,0,0,-q^{n\,\lit}_a)$ such that $Q=\sum_n Q_n$. The positions of the centers, as we discussed in subsection \ref{subsec:Multicenter non-BPS}, are not constrained. The scalar solutions are as in \eqref{zSolelectric} but with the harmonic function of the form:
\begin{equation}
 H_{\lit}=\sum_n \frac{-Q_{n\,\lit}}{\abs{\xb-\xb_n}}+h_{\infty}^{\lit}\,.
\end{equation}
Hence, near the $n$-th center, $z^a$ reads:
\begin{equation}
 z^a=-\im\,\sqrt{\frac{\abs{\varepsilon^{abc}}q_b^{n\,\lit}q_c^{n\,\lit}}{2p^0_{n\,\lit}q_a^{n\,\lit}}}\qquad
\xb\rightarrow\xb_n\,.
\end{equation}
These expressions have the form conjectured in \cite{Kallosh:2006ib}.

We close this section with a remark that our framework admits also the interpretation \cite{Gimon:2009gk} of a non-supersymmetric $stu$ black hole as comprised of supersymmetric constituents. This model follows from the observation \cite{Gimon:2007mh} that the ADM mass
\begin{equation}
 m_{\text{ADM}}=\lim_{\tau\rightarrow 0}\frac{\de U}{\de\tau}\,,
\end{equation}
of a non-BPS black hole can be written as the sum of the masses of four primitive BPS centers. A direct computation in our setting ($\R z=0\Rightarrow B=0$) gives for a non-BPS black hole of electric charge $Q_{\lit}=(p^0_{\lit},0,0,q_a^{\lit})$:
\begin{equation}
 m_{\text{non-BPS}}=k\left(p_{\lit}^0+q^{\lit}_1+q^{\lit}_2+q^{\lit}_3\right),
\end{equation}
with $k$ a constant factor and  $p_{\lit}^0>0$. Computing instead the sum of the masses of four BPS black hole carrying a single type of charge we obtain:
\begin{equation}
 m_{\text{BPS}}=k\left(\lvert p_{\lit}^0\rvert+q^{\lit}_1+q^{\lit}_2+q^{\lit}_3\right).
\end{equation}

Naively, one could try to construct a non-supersymmetric configuration with supersymmetric constituents by taking the matrices $S_n$ to be proportional to the unit matrix. This, however, would not satisfy the condition \eqref{SConstraint}. A way to have supersymmetric centers is to relax the condition of existence of a regular black hole at each of the centers and to assign to each of them only one type of charge. The supersymmetry of such singular configurations will be unaffected by the matrices $S_n$, which we now need to choose equal to the matrix $S$: their effect will be reduced to multiplication by a constant factor.

\section{Conclusions}\label{sec:Conclusions}

In this paper we extended Denef's formalism for multicenter black hole solutions in four-dimensional $\cN=2$ supergravity on simple non-supersymmetric cases, using the fake superpotential method.

Our generalization requires the superpotential to be related to the central charge in a particular way (through a constant matrix $S$), which imposes some constraints on the charge configuration. It turns out to be a limitation, since already some single-centered cases for which the superpotential is known to exist would violate this assumption (cf.~\cite{Ceresole:2007wx}). To satisfy it, we worked with electric or magnetic configurations, which lead to mutually local electric or magnetic constituents.

Still, in the example of the $stu$ model, for the single-centered case we recover the non-supersymmetric black holes previously derived in a different way by Tripathy and Trivedi \cite{Tripathy:2005qp}. The multicenter non-supersymmetric $stu$ solutions that we find, apart from the constraints, correspond to the form conjectured by Kallosh, Sivanandam and Soroush \cite{Kallosh:2006ib}. Our approach allows also to resolve a single non-supersymmetric $stu$ black hole into a collection of supersymmetric centers in a way consistent with the BPS-constituent model of Gimon, Larsen and Sim\'on \cite{Gimon:2009gk}.

More generally, the multicenter solutions that can be described by the method presented here are in a sense the simplest analogues of their supersymmetric counterparts,\footnote{For instance, the Hessian of the black hole potential at its critical points is still proportional to the K\"ahler metric, implying stability and the absence of flat directions.} yet exhibit different properties. In particular, similarly to non-supersymmetric solutions obtained by Gaiotto, Li and Padi \cite{Gaiotto:2007ag} in the group-theoretical approach, but unlike in the generic supersymmetric case, the charges carried by the centers are mutually local and the angular momentum vanishes, rendering the solution static.

The following picture therefore seems to emerge, at least in the considered class of theories: supersymmetric black holes can be split only into supersymmetric composites and only at particular loci of their moduli space, namely on the walls of marginal stability (except for a decomposition into constituents with aligned charge vectors, which is always possible). Non-supersymmetric black holes, on the contrary, can be resolved everywhere in moduli space into a composite consisting of any number of non-supersymmetric centers at arbitrary positions, but also (as Gimon, Larsen and Sim\'on demonstrated for the $stu$ model) into a specific number of threshold-bound supersymmetric constituents (by combining the two descriptions, mixed cases would appear to be also possible).

We are aware, however, that the above summary is not complete. The results of Bena et al.~\cite{Bena:2009ev,Bena:2009en} obtained in the Goldstein and Katmadas's almost-BPS framework \cite{Goldstein:2008fq} demonstrate that non-supersymmetric composites may also comprise constituents with constrained positions. It would therefore be natural to see how the restriction of our method (specifically, the constancy of $S$) could be relaxed, and whether one would then obtain solutions with non-trivial angular momentum. Even more interesting, perhaps, would be to clarify the relationship between the various approaches employed to construct non-supersymmetric multicenter solutions (along the lines of \cite{Bossard:2009we}, for instance, where the superpotential for single-center black holes was obtained through timelike dimensional reduction) and find out whether any of the techniques or their refinements could eventually exhaust all possible classes of extremal solutions. A further step could be then an attempt to use them for non-extremal composites \cite{Stelea:2009ur}.

\acknowledgments

We are grateful to Prof.~Joan Sim\'on for helpful discussions and remarks on a draft of this paper, as well as to Prof.~Antoine Van Proeyen and Bert Vercnocke for valuable conversations. P.G. thanks K.U.Leuven for hospitality. This work is supported in part by the FWO-Vlaanderen, project G.0235.05 and in part by the Federal Office for Scientific, Technical and Cultural Affairs through the `Interuniversity Attraction Poles Programme---Belgian Science
Policy' P6/11-P.

\bibliographystyle{utphys}
\bibliography{Bibliography}

\end{document}